\documentclass[conference]{IEEEtran}

\IEEEoverridecommandlockouts
\usepackage{amsmath}
\usepackage{color}
\usepackage{array}
\usepackage{stfloats}
\usepackage{amsthm} 
\usepackage{amssymb}
\usepackage{float}
\usepackage{nccmath}
\usepackage{graphicx}
\usepackage[linesnumbered,ruled]{algorithm2e}
\usepackage{setspace}
\usepackage{booktabs}
\usepackage{enumitem}
\usepackage{subcaption}
\usepackage{mwe}
\usepackage{cite}
\usepackage{amsmath,amssymb,amsfonts}
\usepackage{graphicx}
\usepackage{tabularx}
\usepackage{textcomp}
\usepackage{xcolor}
\usepackage[utf8]{inputenc} 
\usepackage{kotex} 

\usepackage{algorithmicx}
\usepackage{xspace}
\def\BibTeX{{\rm B\kern-.05em{\sc i\kern-.025em b}\kern-.08em
    T\kern-.1667em\lower.7ex\hbox{E}\kern-.125emX}}
\begin{document}
\newcommand{\BfPara}[1]{{\noindent\bf#1.}\xspace}
\newcommand\mycaption[2]{\caption{#1\newline\small#2}}
\newcommand\mycap[3]{\caption{#1\newline\small#2\newline\small#3}}

\title{The Credibility Cryptocurrency Valuation: Statistical Learning Analysis for Influencer Tweets}

\author{
\IEEEauthorblockN{Haemin Lee}
\IEEEauthorblockA{\textit{Korea Univ.}\\
haemin2@korea.ac.kr}
\and
\IEEEauthorblockN{Youngkee Kim}
\IEEEauthorblockA{\textit{Korea Univ.}
\\
felixkim@korea.ac.kr}
\and
\IEEEauthorblockN{Hyunhee Cho}
\IEEEauthorblockA{\textit{Sungkyunkwan Univ.}\\
gt7746d@g.skku.edu}
\and
\IEEEauthorblockN{Soyi Jung}
\IEEEauthorblockA{\textit{Hallym Univ.}\\
sjung@hallym.ac.kr}
\and
\IEEEauthorblockN{Joongheon Kim}
\IEEEauthorblockA{\textit{Korea Univ.}\\
joongheon@korea.ac.kr}
}

\maketitle

\begin{abstract}
Cryptocurrency has attracted significant attention. Considering the number of individuals investing in bitcoin, their motivations are comparatively less clear than traditional investment decisions. As of December 2020, the market has continuously increased in cryptocurrency. Especially, the spike of joke Dogecoin shows the weirdness of the modern meme economy with the support of Elon Musk, whom himself appointed as "Dogefather".
In this paper, we analysis the impact of tweets by Elon musk and present some statistical analyze with event study.
\end{abstract}


\section{Introduction}

Cryptocurrency has attracted great attention to speculators and investors. Apart from its popularity, with its hyper volatile characteristics(i.e., skyrocketing and sudden fall), it is hard to make investment decisions or predict the price fluctuation. As a sequence, there are many studies on cryptocurrency prediction with financial time-series data~\cite{isj202003saad, isj202103kim,bc1,bc2,bc3,bc4,bc5,bc6,bc7,bc8,9333853,9333933}.
However, the motivations of many individuals who invest in bitcoin are unclear and greatly influenced by media or regulations compared to the traditional stock market. Especially, Dogecoin has skyrocketed 1200\% since January 2021, showing abnormal behavior. Therefore, it is necessary to investigate the abnormal aspects of Dogecoin.

Dogecoin is a virtual currency based on the famous Internet meme "Shibe doge." The coin makers Billy Markus and Jackson Palmer reportedly launched the Dogecoin with the logo of a Shiba dog in hopes of the coin becoming a loved and entertaining cryptocurrency. Tesla CEO Elon Musk has shown his love for Dogecoin on Twitter several times. Dogecoin is a currency made as a joke to satirize the Bitcoin frenzy in the coin market. However, it is difficult to explain the abnormal price increase other than the Musk effect.


Now, investors sell Dogecoin as soon as the price rises, hold on in the coin market, or buy and sell coins affected by the mention of Dodgecoin on Elon Musk's Twitter account. Therefore, there are many studies on the connection between Dogecoin and Musk tweets~\cite{ante2021elon}. Among those researches, we'd like to investigate the relationship between "Dogecoin Image Tweet of Elon Musk" and stock price volatility.

\begin{figure}
    \centering
    \includegraphics[width=1.0\columnwidth]{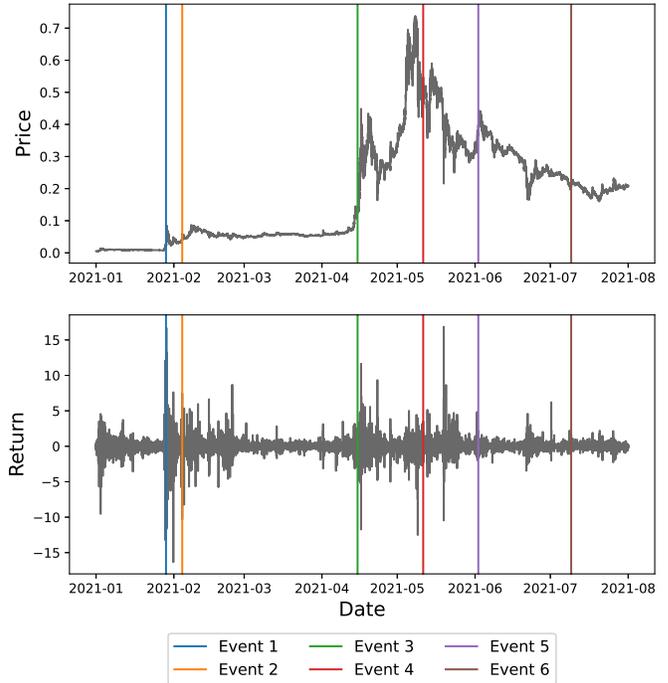}
    \caption{Dogecoin's Year-to-Date Performance}
    \label{fig:performance}
\end{figure}

Figure~\ref{fig:performance} shows the year-to-date performance trend of Dogecoin. The vertical lines are drawn when Musk's twit event occurs. It can be seen that the movement of price and return values change from every six vertical lines. In this study, we assume Elon Musk's tweets affect cryptocurrency and examine whether abnormal price trends appear only as Dogecoin and how they affect other mainstream coins.

The objectives of the study are to statistically infer Elon musk's influence on the Dogecoin market and, if it is, to find out whether Musk's impact decreases over time or continues as he keeps posting the tweets based on the shepherd's boy effect.

\section{Data and Methods}

\begin{table*}[t]
\small
\begin{center}
\begin{tabular}{ccp{0.5\textwidth}}
\toprule
Event &  Date & Content  \\ 
\midrule
1 & 2021-01-28 20:47 & \begin{itemize}
      \item Posted the cover of a magazine named Dogue.
      \end{itemize}
      \\
\midrule
2 & 2021-02-04 07:57 & \begin{itemize}
      \item Tweeted $\colon$ dogecoin
      \end{itemize}
      \\
\midrule
3 & 2021-04-15 4:33 & \begin{itemize}
      \item Tweeted $\colon$ Doge Barking at the Moon.
      \item Posted the picture of painting named "Dog Barking at the Moon (1926)" by Spanish artist Joan Miró.
      \end{itemize}
      \\
\midrule
4 & 2021-05-11 8:13 &  \begin{itemize}
      \item Tweeted $\colon$ Do you want Tesla to accept Doge?
      \item Posted the vote whether Tesla to accept Dogecoin.
      \end{itemize}
      \\
\midrule
5 & 2021-06-02 07:05 &  \begin{itemize}
      \item Found this pic of me as a child
      \item Posted the picture of Doge with the caption saying "I have to keep my passion hidden from the public or I'll be socially ostracized".
      \end{itemize}
      \\
\midrule
6 & 2021-07-09 7:15 &  \begin{itemize}
      \item Retweeted the post of Matt Wallace $\colon$ Merit of Doge comparing with BTC \& ETH.
      \end{itemize}
      \\
\bottomrule
\end{tabular}
\caption{Doge-related events on Elon Musk's Twit}
\label{tbl:event}
\end{center}
\end{table*}

The Cryptocurrency data of BTC/USDT, DOGE/USDT used in this paper is collected through BINANCE API from 2021.01 to 2021.07. We used the minute OHLC(i.e., open, high, low, closing prices) data for each minute and additionally calculated the log return, which is calculated by taking the natural log of the previous closing value divided by the closing value. The one-period log return is given as $R_{t} = \ln{P_{t}} - \ln{P_{t-1}}$. 
%
For the Elon Musk tweet data, we chose the six events from January to July and checked out their impacts. The contents of tweet events are summarized in Table~\ref{tbl:event}.







\begin{figure*}
\centering
\subfloat[Event 1]{\includegraphics[width=0.99\columnwidth]{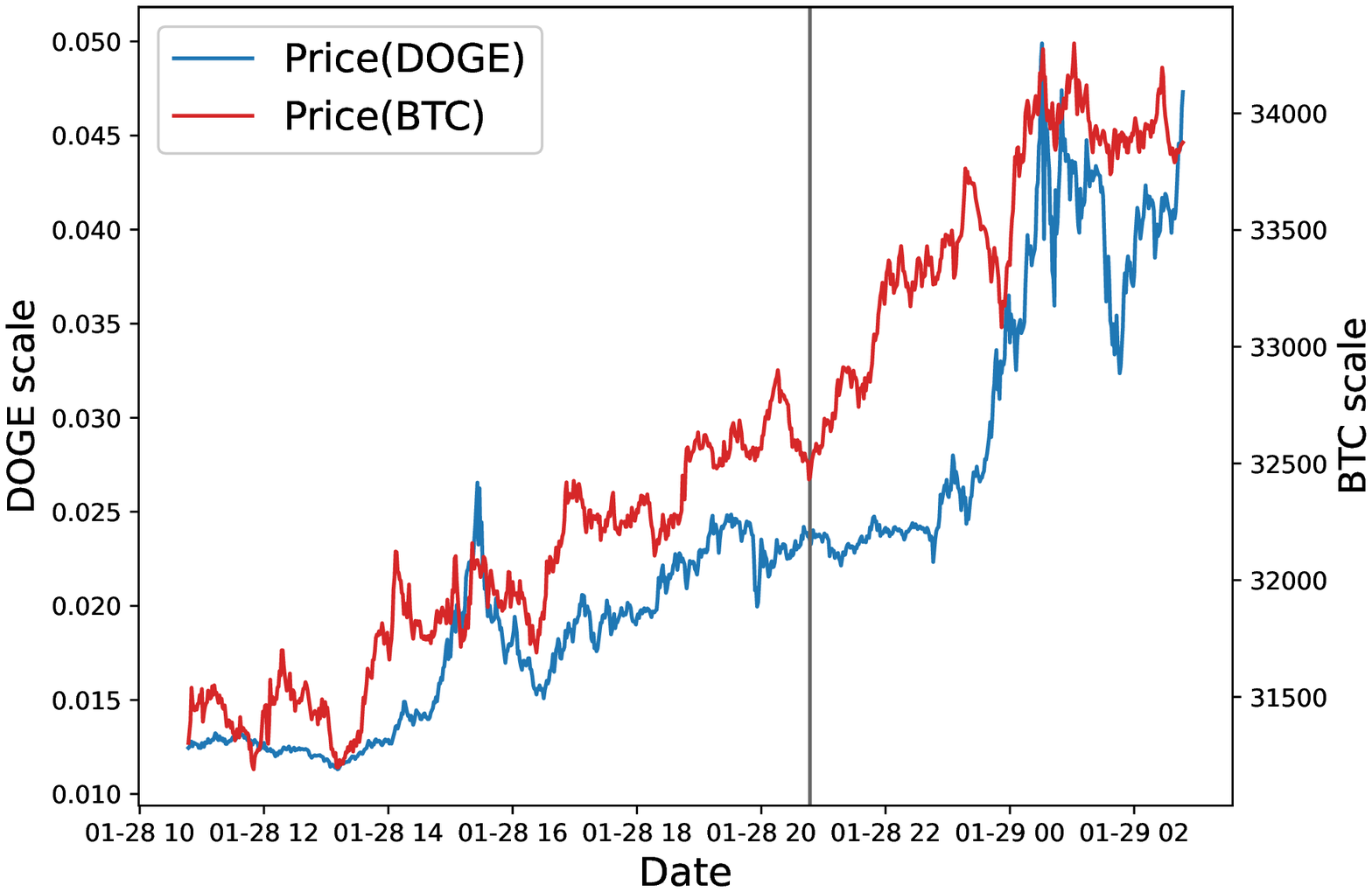}}
\subfloat[Event 2]{\includegraphics[width=0.99\columnwidth]{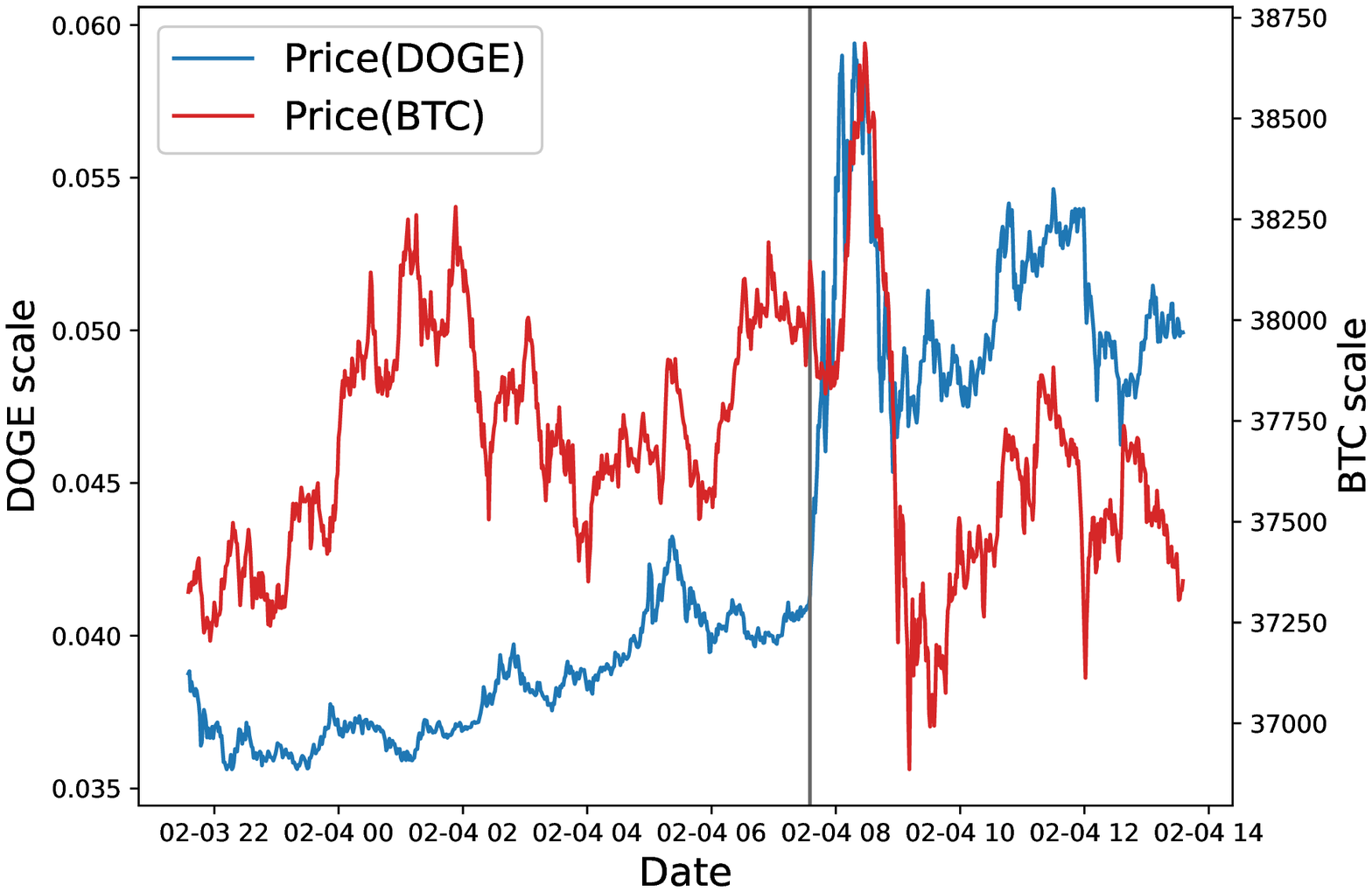}} 
\hfil
\subfloat[Event 3]{\includegraphics[width=0.99\columnwidth]{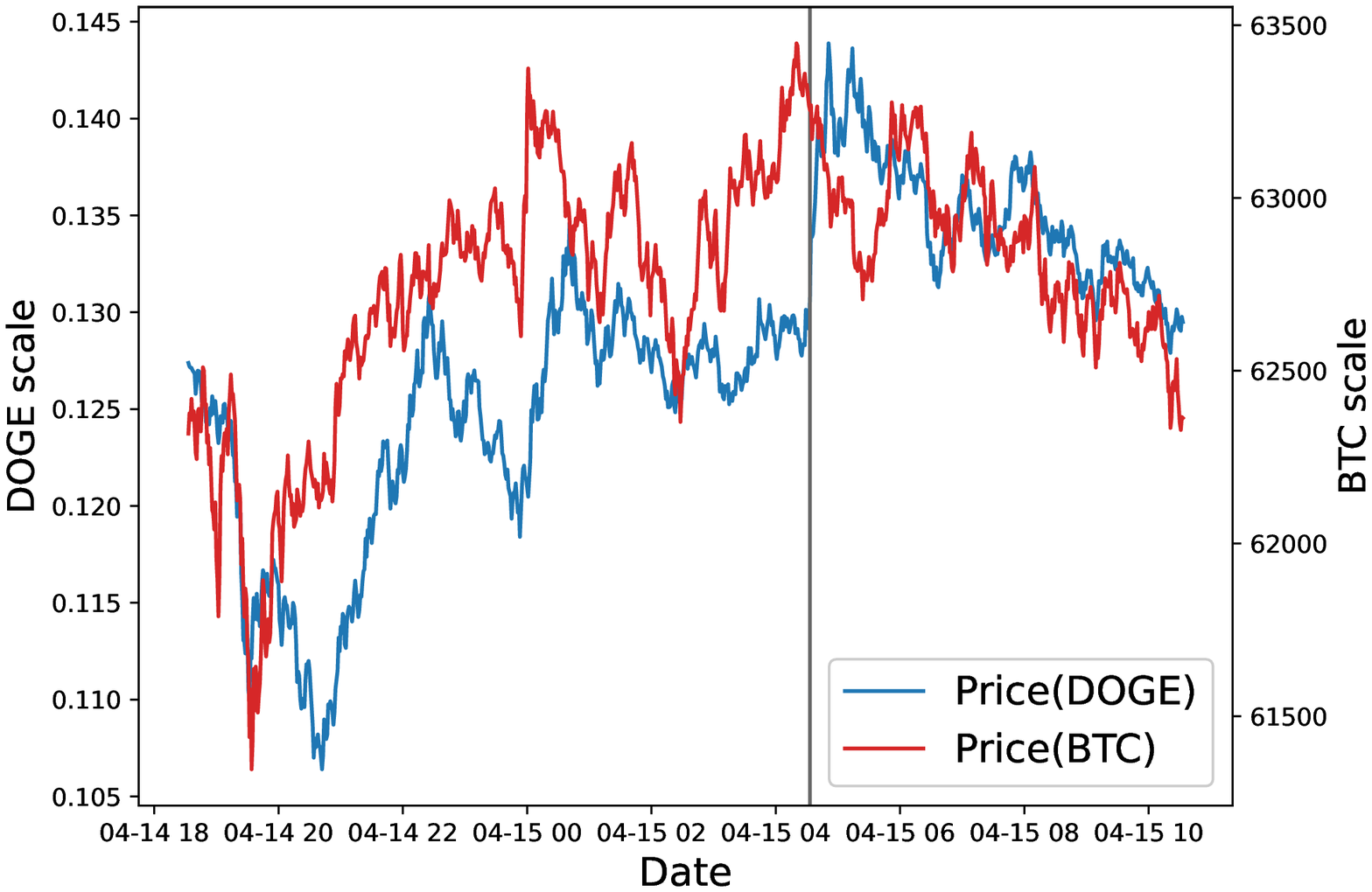}}
\subfloat[Event 4]{\includegraphics[width=0.99\columnwidth]{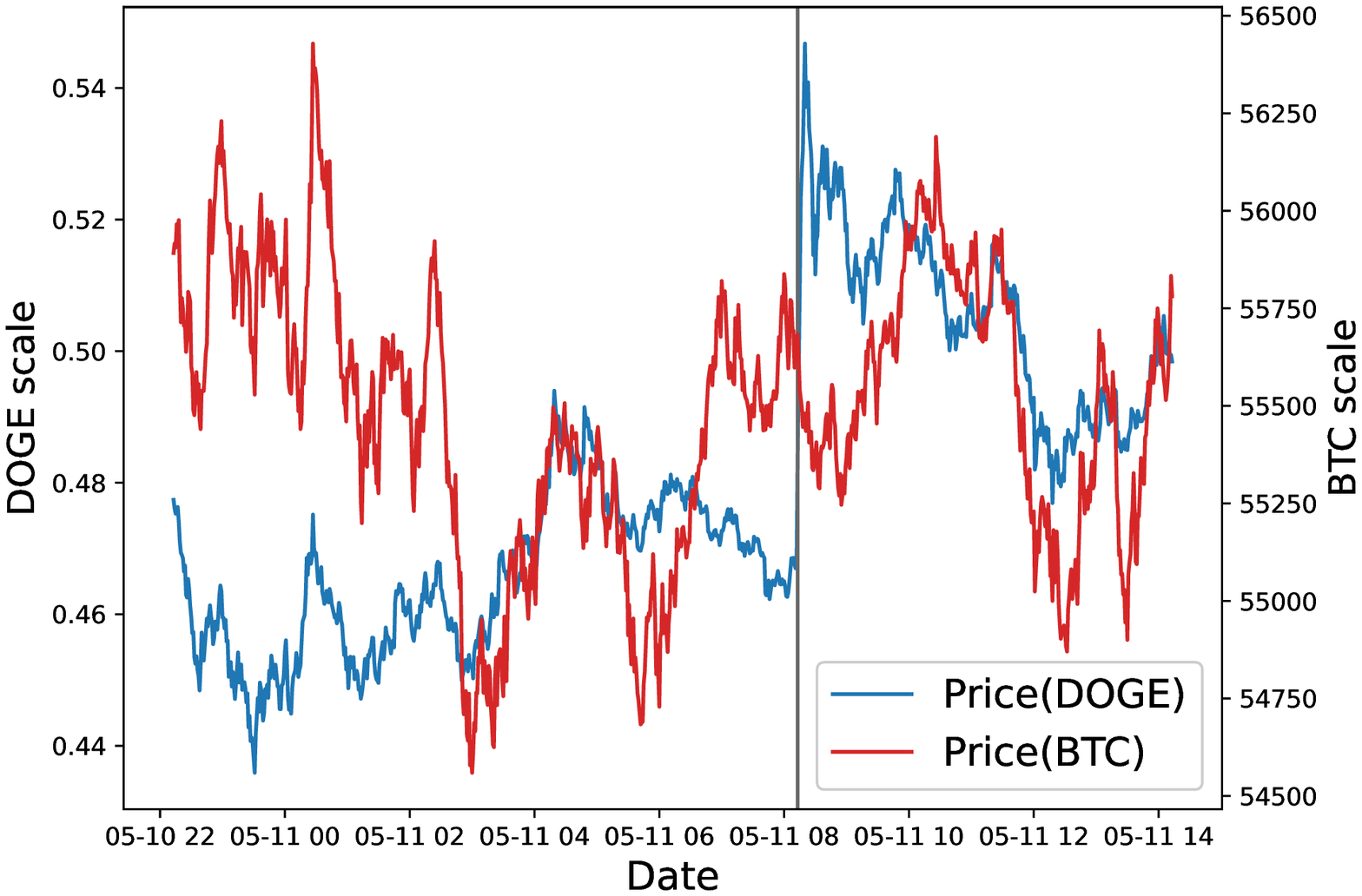}}
\hfil
\subfloat[Event 5]{\includegraphics[width=0.99\columnwidth]{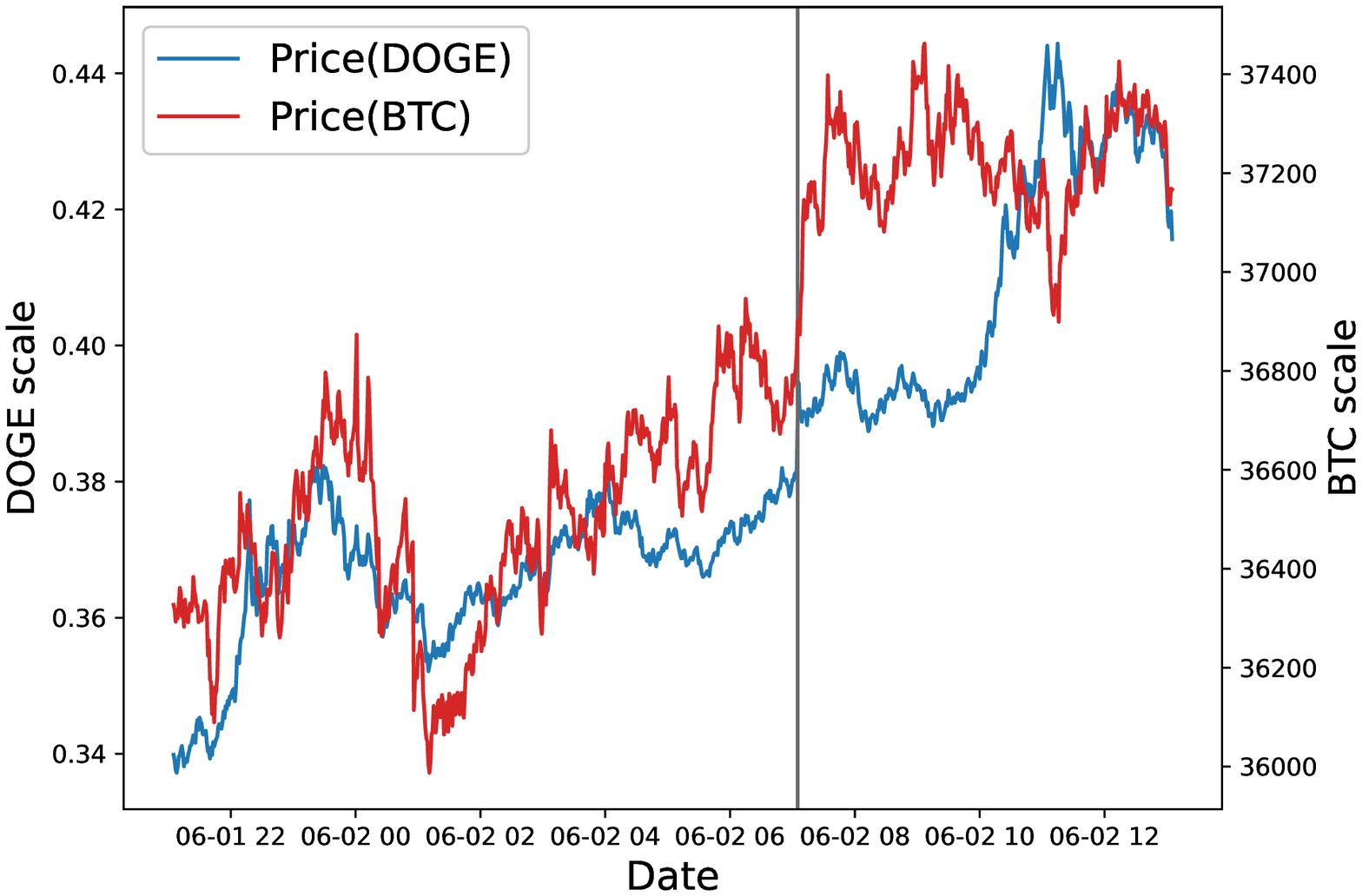}}
\subfloat[Event 6]{\includegraphics[width=0.99\columnwidth]{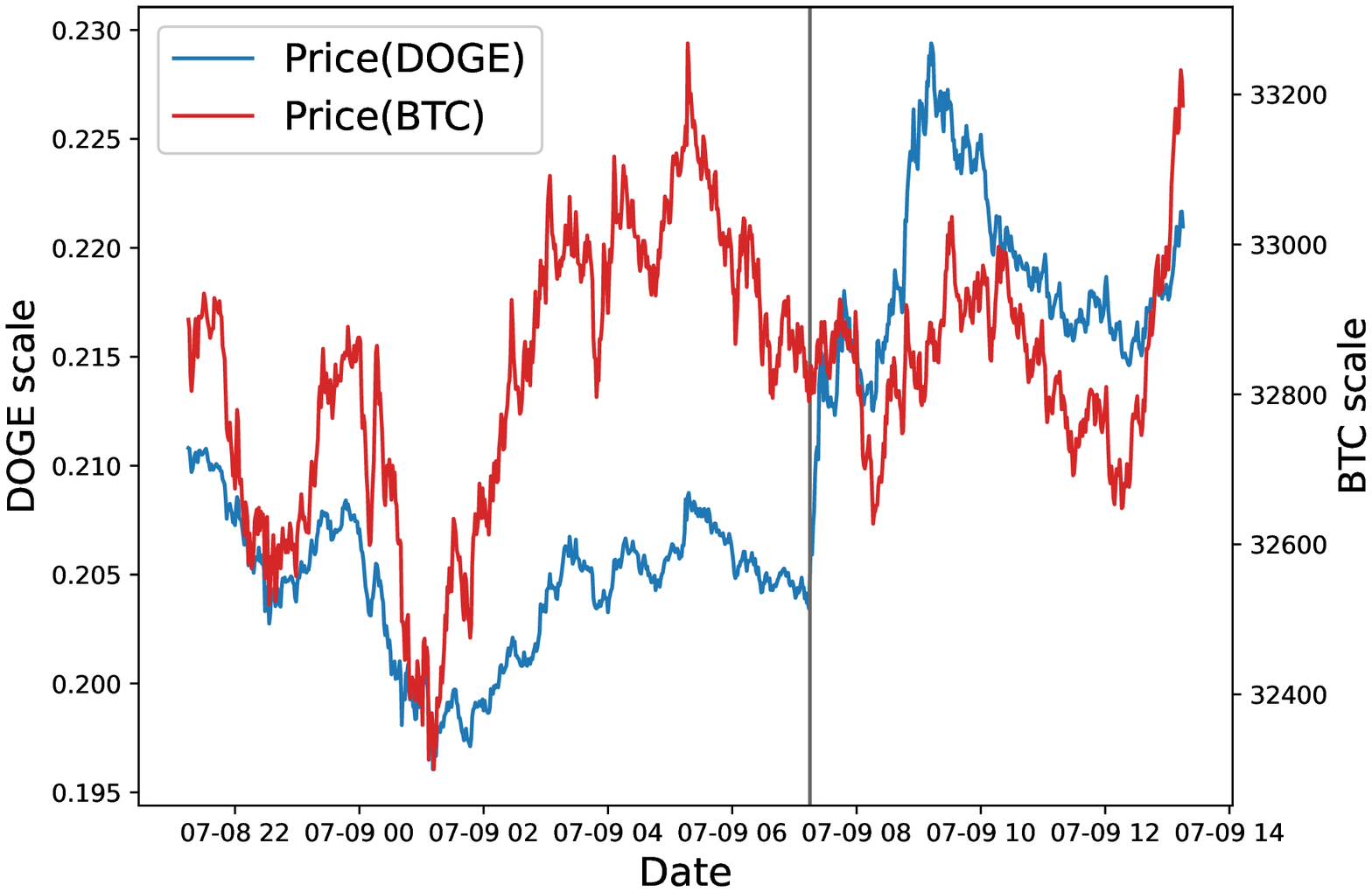}}
\caption{Price of Dogecoin and Bitcoin for events}\label{fig:price_event}
\end{figure*}

Figure~\ref{fig:performance} shows Dogecoin's year-to-date price and return value.
In most events, it can be seen that the price had soared when the event occurred.
Additionally, Fig~\ref{fig:price_event} shows the price flow of BTC/USDT and DOGE/USDT for the individual events.
The vertical gray line indicates the time of Elon Musk's tweet. We can glean the dogecoin rise after the tweet event. Also, we can roughly grasp the price correlation between DOGE/USDT and BTC/USDT, considering the two time-series have a different price range.


%
For the event study, since the tweet's effect may disappear quickly, we only take the short interval after the event. We set the event window as 10 hours before and after the event time ($-10 \sim 10$, $t= 0$) to measure the impact over a particular time period.
The most important part of the event study is the calculation of abnormal returns due to the advent of certain unexpected events~\cite{brown1985using}. Abnormal returns can be estimated at normal and actual return rates estimated from the prediction model in the estimation window.
For the calculation of expected return, we used the constant mean return model (Brown and Warner, 1985) which is frequently used in economics. The model calculates the mean return over a 9 hour time period before the event(t = -600 to -60 minutes), which is sufficient length~\cite{armitage1995event}.
For the abnormal return calculation, we took a static approach setting a 10\% threshold. Bremer and Sweeney (1991) use a 10\% price change as a criterion for abnormal returns, which is suitable in the case of the cryptocurrency markets~\cite{bremer1991reversal}. Subsequently, the cumulative abnormal return (CAR) can be obtained by the sum of all the abnormal returns used in the period as in~\eqref{eq: CAR}.

\begin{equation}
    CAR_{(t_1,t_2)} = \sum_{i = 1}^{100}{AR_{i,t}}
\label{eq: CAR}
\end{equation}

In order to determine whether the tweet actually affected the stock price, we have to verify that the null hypothesis that this value is 0 can be statistically rejected. In general, when the null hypothesis can be rejected at the significance level of \,5\%, it can be inferred that the price of Dogecoin is affected by the events.
We conduct the two-tailed one-sample t-test to compare the means of predicted return and true return, that is, to verify whether the CAR is zero. The null hypothesis ($H_0$) is that the true difference between is zero. The alternate hypothesis ($H_a$) is that the true difference is different from zero. The t-statistic is as following~\eqref{eq: tstat}.

\begin{equation}
\begin{aligned}
    S_{CAR} = & \sqrt{n \times MSE} = \sqrt{n \times \frac{\sum_{t = 1}^{m}{\hat{\epsilon}_{t}^2}}{m-2}}, \\
    t = & \frac{CAR_{t}}{S_{CAR}}
    \label{eq: tstat}
\end{aligned}
\end{equation}

\section{Results}

The price and return value apparently seem to be related to Elon Musk's twit events, as shown in Figure~\ref{fig:performance}. However, due to the hypervariable nature of cryptocurrency, it is difficult to grasp the real impacts correlated with 
Musk's tweet. In this section, we took a statistical inference to assess the impact of each event. We first assumed that there are no other factors rather than Musk's tweet that would affect the price of Dogecoin. Considering its position in the cryptocurrency market as prank coin, our assumption seems reasonable, and we apply an event study taking Musk's tweets as individual events to verify the real price impacts by Musk.
Each column in table~\ref{tab:car} refers to the six selected events and shows the threshold applied CAR and t-statistic for every event over different time periods from 5 minutes to 360 minutes. Each *, **, and *** indicates significance at the level of 10\,\% , 5\,\%, and  1\,\%.
We can infer that events 1, 3, 4, and 6 have some significant statistics With the hypothesis testing results. For the case of 3 and 4, CAR rise to 7.10\,\% and 14.48\,\% after 10 minutes. For event 6, the effect lasted for 120 minutes, reaching CAR to 12.07\,\%.


\begin{table*}
\centering
\resizebox{\textwidth}{!}{
\begin{tabular}{lccccccccccccc}
\toprule
Event & \multicolumn{2}{c}{event 1} & \multicolumn{2}{c}{event 2} & \multicolumn{2}{c}{event 3} & \multicolumn{2}{c}{event 4} & \multicolumn{2}{c}{event 5} & \multicolumn{2}{c}{event 6}\\
\cmidrule(lr){2-3} \cmidrule(lr){4-5} \cmidrule(lr){6-7} \cmidrule(lr){8-9} \cmidrule(lr){10-11} \cmidrule(lr){12-13}
Window & CAR & t-statistic & CAR  & t-statistic  & CAR  & t-statistic & CAR & t-statistic & CAR & t-statistic & CAR & t-statistic \\
\midrule
0 to 5  & -1.01\,\% & -0.31 & 7.02\,\% &2.29* & 4.95\,\% & 2.18* & 12.56\,\% & 4.42** & 2.35\,\% & 0.78 & 3.02\,\% & 2.72** \\
0 to 10 & -1.53\,\% & -0.47 & 14.67\,\% & 3.89*** & 7.10\,\% & 2.48** & 14.48\,\% & 3.05** & 2.51\,\% & 0.82 & 5.34\,\% & 2.89** \\
0 to 30 & -10.67\,\% & -1.74* & 35.10\,\% & 2.12** & 7.79\,\% & 1.63 & 10.95\,\% & 1.45 & 3.48\,\% & 1.09 & 5.79\,\% & 2.03* \\
0 to 60 & -4.68\,\% & -0.58 & 27.68\,\% & 1.23 & 7.19\,\% & 1.23 & 8.82\,\% & 1.09 & 1.86\,\% & 0.51 & 4.64\,\% & 1.35 \\
0 to 120 & -18.28\,\% & -1.58 & 15.99\,\% & 0.60 & 2.65\,\% & 0.39 & 9.04\,\% & 1.04 & 1.49\,\% & 0.38 & 12.07\,\% & 2.68** \\
0 to 180 & 12.20\,\% & 0.60 & 19.72\,\% & 0.72 & 3.55\,\% & 0.47 & 7.47\,\% & 0.82 & 1.68\,\% & 0.38 & 9.85\,\% & 1.95* \\
0 to 240 & 27.16\,\% & 0.62 & 24.62\,\% & 0.86 & 4.07\,\% & 0.51 & 3.78\,\% & 0.38 & 12.02\,\% & 2.08 & 8.50\,\% & 1.60 \\
0 to 300 & -5.16\,\% & -0.10 & 9.87\,\% & 0.33 & 4.13\,\% & 0.48 & 4.45\,\% & 0.42 & 8.11\,\% & 1.13 & 7.33\,\% & 1.32 \\
0 to 360 & 24.25\,\% & 0.47 & 17.14\,\% & 0.56 & 1.73\,\% & 0.19 & 6.45\,\% & 0.59 & 3.71\,\% & 0.47 & 9.72\,\% & 1.69 \\
\bottomrule
\end{tabular}}
\caption{Cumulative Abnormal Return}
\label{tab:car}
\end{table*}

\section{Concluding remarks}
In this paper, we give a short analysis between Elon Musk's tweet and Dogecoin's price valuation. For deeper analysis, the event study result is presented. We applied 10\,\% price threshold for CAR calculation and showed the statistical significance of Musk's tweet events.
However, we came with the assumption that there is no other significant factor that would affect the Dogecoin's price without the tweets. In the real world, there might be some other events that need to be considered. The cryptocurrency market sentiment and political issues might affect the price. 
Due to the hyper-variant and unstable nature of cryptocurrency, it is hard to derive an accurate casual effect.
Moreover, a more sophisticated approach to financial time-series data would bring some insights. The configuring of a predicting model and event selection can be considered as future work analysis. Finding some numerical inference on whether a cryptocurrency will eventually be immune to Elon Musk's tweet would be interesting.


\section*{Acknowledgment}
This research is supported by National Research Foundation of Korea (NRF-Korea 2019R1A2C4070663). J. Kim is the corresponding author of this paper.

\bibliographystyle{IEEEtran}
\bibliography{ref_aimlab, bitcoin}

\end{document}